\documentclass[aps,prd,twocolumn,showpacs,groupedaddress,nofootinbib]{revtex4}
\usepackage{graphicx}
\begin{document}

\title{Phase diagram of QCD with 2+1 flavors of Wilson quarks at finite temperature and chemical potential}

\author{He-Sheng Chen}
\thanks{Corresponding author. Email address: chenjs@yzu.edu.cn}
\address{Department of Physical Science and Technology, Yangzhou University, Yangzhou 225009, China}

\author{Xiang-Qian Luo}
\address{
CCAST (World Laboratory), P.O. Box 8730,
Beijing 100080, China\\
and Department of Physics, Zhongshan (Sun Yat-Sen) University, Guangzhou 510275, China}
\thanks{Mailing address.}

\date{\today}

\begin{abstract}
The first results on lattice at finite temperature $T$ and chemical potential $\mu$ with 2+1
flavors of Wilson quarks are presented. We have studied the dependence of chiral phase transition
and deconfinement phase transition on quark mass. Finite volume size analysis and Binder cumulants
are used to determine the properties of phase transition. Phase diagram of QCD with 2+1 flavors of
Wilson quarks are presented.
\end{abstract}


\maketitle

\section{Introduction}
   With the heavy-ion collision experiments at RHIC now running, it is much more important to understand
the phase diagram of QCD at finite temperature and finite density. QCD predicts the existence of
quark-gluon plasma(QGP) at high temperature and the existence of phase transition between the
hadronic matter and the QGP. But there is no direct evidence yet. It is a very important work to
locate the phase transition region and to investigate its properties.

   The lattice gauge theory (LGT) is the unique nonperturbative method for the quantum gauge
field. Lots of MC(Monte carlo) simulations for QCD on lattice were performed to study the quark
mass spectrum,gluon ball mass spectrum,phase diagram and so on.

In the past, the most MC simulations for QCD at finite temperature and finite density were
performed with staggered quarks owing to its preserving the continuous $U(1)$ chiral symmetry .
But staggered fermions formalism on the lattice does not completely solved the species doubling
problem. For to reduce the fermionic degrees of freedom, MC simulation take the fourth root of the
fermion determinant to thin the species doubling problem. The operation can arise the locality
problem in simulations with non 4 taste fermions \cite{Neuberger:2004be}. The Wilson fermion
formalism preserves the flavor symmetry and has no the species doubling problem. It is also only
known as a formalism that possesses a local action for any number of flavors and is well formalism
for MC simulation on lattice.

   In our previous work{\cite{hschen:2005mq}}, the first phase diagram of QCD with 4 flavors($N_f=4$)
of Wilson quarks at small chemical potential and high temperature was presented by using imaginary
chemical potential method. The critical line of first order phase transition was located at
$(\mu,T)$ plane. $N_f=4$ presume that there are four degenerate quarks. But the physically
relevant case should be two degenerate massless u,d quarks and one light S quark, which is
so-called 2+1 flavors quarks. It is important to investigate the phase diagram of QCD with 2+1
flavor quarks for us to understand the actual state of high density star and the cosmic origin.

In this paper, we will present our results of MC simulations with 2+1 flavors of Wilson quarks at
finite chemical potential and finite temperature by using the pure imaginary chemical potential
method.

In section II, we will briefly instruct the latterly progress in this domain. In section III, we
will presented the formalism used in this paper. In section IV, our detailed results of MC
simulation were presented. At last, we will present the conclusion.

\section{General characteristic of the phase transition for QCD}

  \begin{figure} [htbp]
  \begin{center}
  \includegraphics[totalheight=2.0in]{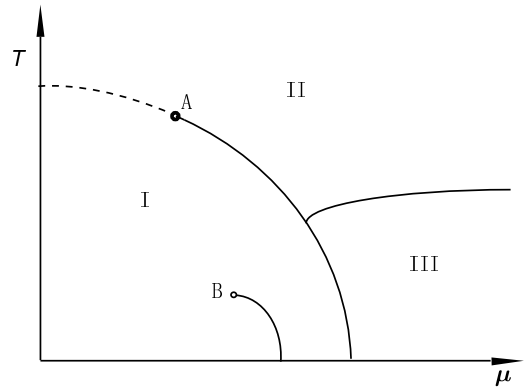}
  \end{center}
  \caption{The schematic phase diagram of QCD in ($\mu,T$) plane} \label{fig-1}
  \end{figure}

  Before starting our discussion, let us briefly review present understanding of the phase diagram structure.
  The phase diagram in the ($\mu,T$) Plane is shown in Fig.\ref{fig-1}. The ($\mu,T$) Plane
  is divided into three main parts, I, II, III. At I domain, where the temperature and chemical potential are
  below critical line, the QCD matter was formed as hadrons and the quarks are confined. At the II domain,
  because of extra high temperature and strong interacting, quarks are released and the strongly interacting
  QCD matter formed the quark-gluon plasma(QGP) and the quarks are deconfined. In the III domain, where the
  temperature is low and the density of quarks is very high, it is believe that two quarks on the fermion surface
  can be bound into an Cooper pair, which is so-called color superconductivity. That case is similar to the
  superconductivity in the electromagnetism. The solid line which end at the point A is the the first order
  critical line. The dotted line which start from the point A is the the second order critical line.
  The point A is the trecritical point. It is very important to accurately locate the trecritical point for searching and
  investigating the QGP.

  \begin{figure} [htbp]
  \begin{center}
  \includegraphics[totalheight=2.5in]{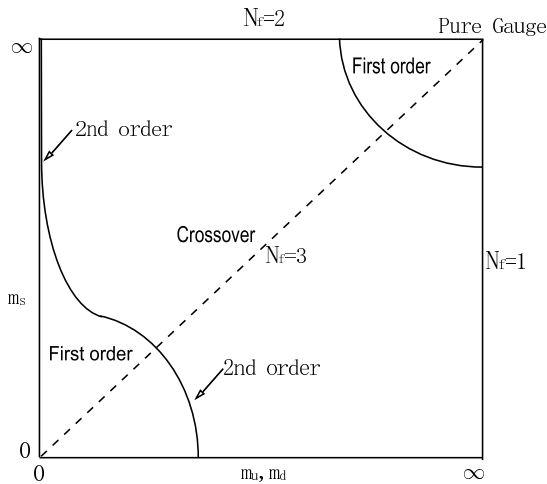}
  \end{center}
  \caption{The schematic phase diagram of QCD in ($m_{ud},m_s$) plane for $\mu=0$} \label{fig-2}
  \end{figure}

  According to the MC simulation result of the past on lattice, the nature of phase transition of QCD depends on
  the quantity of quark mass. At zero chemical potential $\mu = 0$, the schematic phase diagram is shown in
  Fig.\ref{fig-2} in the ($m_{ud}$,$m_s$) plane.
  For the pure gauge case, for which all the quark mass degenerate into infinitely large, a first order transition
  is expected in the quenched lattice gauge theory.
  For QCD with two degenerated light u,d quarks and a infinitely large s quark, which is so-called two flavors QCD,
  it is suggested that the chiral transition is of second order in the limit of vanishing quark mass\cite{Karsch:1994hm}.
  For three or more degenerate quarks, the transition is characterized as first order while the quantity of quark
  mass is less than some critical value $m^c_q$\cite{deForcrand:2003hx}. For the non-degenerate quarks, the strange
  quark is heavier than the u,d quark, one expects a critical line which separates the regime of first order phase
  transitions from the crossover regime\cite{Schmidt:2002uk}. In the other area, the transition is
  continuous. The structure of phase diagram is sensitive to the the non-degenerated quark mass.

\section{The Formalism in Lattice Gauge Theory}
In the thermal field theory, the canonical partition function $Z(V,T)$ can be expressed as:
       \begin{equation}
        \label{QCD_partition}
          Z= \int [dU][d\bar\psi][d\psi]e^{-( S_G(U)+S_F(U,\bar\psi,\psi))},
       \end{equation}
where $S_G$ is the gauge field action:
$$
  S_G=\frac{1}{2}\int_0^{1/T}d\tau\int_v d^3\vec{x} F_{\mu\nu}F_{\mu\nu},
$$
$S_F$ is the fermion field action:
$$
  S_f=\int_0^{1/T}d\tau\int_v d^3\vec{x}\sum_{f=1}^{N_f}{\bar\psi}_f (\gamma_\mu D_\mu+ m_f -\mu\gamma_4) \psi_f,
$$
where $m_f$ is the quark mass.

On lattice, the canonical partition function $Z$ is:
\begin{eqnarray}
   Z= \int [dU]\prod_{f=1}^{N_f}detM(U,\kappa_f,a\mu)e^{-\beta S_G(U)},
\end{eqnarray}
where, $\kappa_f = 1/(8+2am_f)$ is the quark mass parameter, $a$ is the lattice space, $\mu$ is
chemical potential. The gauge field action $S_G$ is:
\begin{eqnarray}
   S_G(U)  =-\beta \sum_{\mu,\nu<\mu} P_{\mu,\nu},
\end{eqnarray}
where $P_{\mu,\nu}$ is a plaquette on lattice. For QCD with Wilson fermions, the fermion action
$S_F$ is :
\begin{eqnarray}
   S_F(\bar\psi,\psi,U)  = \bar\psi M \psi,
\end{eqnarray}
where $M$ is the fermionic determinant:
\begin{eqnarray}
   M_{i,j} & = & \delta_{i,j}+\kappa_f\{\sum_{i=1}^{3}[(1-\gamma_i)U_\mu(n)+
          (1+\gamma_i)U^\dagger_i(\hat{n}-\hat{i})]\nonumber \nonumber \\
          &   & +[(1-\gamma_4) e^{a\mu}U_4(n)+ (1+\gamma_4)e^{-a\mu}U^\dagger_4(\hat{n}-\hat{4})]\}.\nonumber \\
          &   &
  \label{WS action}
  \end{eqnarray}
For $\mu = 0$, $detM(U,\kappa_f,a\mu)$ is positive real, MC methods can be used to simulate the
thermodynamics of QCD. But for $\mu \neq 0$, $detM(U,\kappa_f,a\mu)$ is complex, the standard MC
simulation methods are invalid. Some improved MC methods can be introduced in simulation of QCD .
In this paper, the pure imaginary chemical potential method is chosen, which is proper for low
baryon density and high temperature.

QCD is a theory of asymptotic freedom. The strength of interaction between two static quarks could
reflects on the state of quarks. Strong interaction indicate that quarks are isolated and
deconfined. Polyakov loop
\begin{eqnarray}
    \label{Polyakovloop}
    P(\vec{x})  =   {\rm Tr} \left[\prod_{t=0}^{N_t -1}{U_4(\vec{x},t)} \right]
\end{eqnarray}
can tell us the strength of interaction between two static quarks, which can be used to determine
the confinement-deconfinement phase transition. The chiral condensate
\begin{eqnarray}
    \label{p_bar_p}
    \langle \bar\psi \psi \rangle &=& \frac{1}{Z}\int [dU] [d\bar\psi][d\psi]\bar\psi \psi e^{-S_g-S_f}
\nonumber \\
             &=&{1\over Z VN_t} \int [dU] {\rm Tr} \left( M^{-1}[U] \right)   \left({\rm Det} M[U]\right)^{N_f} e^{-S_g}  ,
\nonumber \\
\end{eqnarray}
is the order parameter of chiral-symmetry breaking. At the chiral limit, non-zero of $\langle
\bar\psi \psi \rangle$ implied that the chiral-symmetry of QCD is spontaneously broken. According
to the present understanding, the chiral transition point is same as the deconfinement transition
one.

In this paper, several methods are used to locate the critical point of transition. A method is to
search the location of rapid variety of order parameters. The second one is to search the peaks of
the order parameter fluctuations at critical points. It is convenient to determine the
universality class of transition by using the Binder cumulants of physical quantity at the
transition point. The dimensionless fourth order cumulants $B_4$ of an arbitrary observable
physical quantity $x$ is constructed as
\begin{eqnarray}
    \label{B_4}
    B_4(m_c,\mu_c,\beta_c) = \frac{<(\delta x)^4>}{<(\delta x)^2>^2},
\end{eqnarray}
where, $m_c,\mu_c,\beta_c$ is the value at critical point. Binder cumulants calculated on
different size lattices will intersect at some value of the quark mass. At the infinite volume
limit, these intersection points will converge to a universal value which is characteristic for
the universality class. The universal value is quite different for different symmetric system. In
$N_f=3$, the chiral symmetry of QCD at critical point is like 3D Ising model, the intersection
points converge to $B_4 = 1.604$ \cite{Schmidt:2001kq}. In this paper, $B_4$ is used to determine
the characteristic of chiral phase transition.

\section{Simulation results for QCD with 2+1 flavors fermions}
In this paper, MC simulation are performed with the standard $R$ algorithm\cite{Gottlieb:1987mq}.
We have modified the MILC collaboration's public code\cite{Milc} to fit the imaginary chemical
potential and 2+1 flavor case. MC simulations are performed at lattice size $V\times N_t =8^3
\times 4$ and finite size scaling analysis was checked for some different lattice size in certain
parameters. The parameter $\kappa_1$, which corresponds to parameter of $u,d$ quark mass, was
scanned from $0.16$ to $0.20$. The parameter $\kappa_2$, which corresponds the mass parameter of s
quark and is heavier than $u,d$ quark, was fixed at $\kappa_2=0.08$ which is much less than
$\kappa_1$. The $\delta \tau$ is $0.02$. Each configuration runs 20 molecular steps with
microcanonical step size $\delta \tau = 0.02$. A configuration is token for measurement every
other 20 configurations.

\begin{figure} [htbp]
\begin{center}
\includegraphics[totalheight=2.0in]{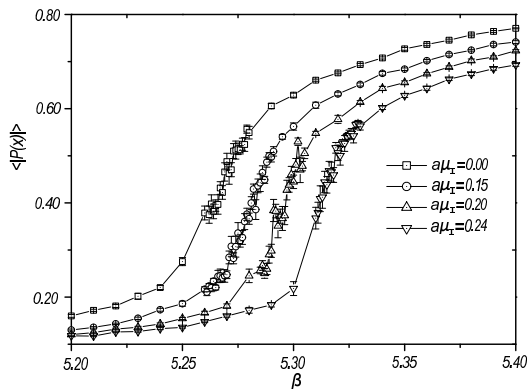}
\end{center}
\caption{Polyakov loop norm as a function of $\beta$ for $a\mu_I=0.0,0.15,0.20,0.24$ in
$\kappa_1=0.16$.} \label{fig-3}
\end{figure}

\begin{figure} [htbp]
\begin{center}
\includegraphics[totalheight=2.0in]{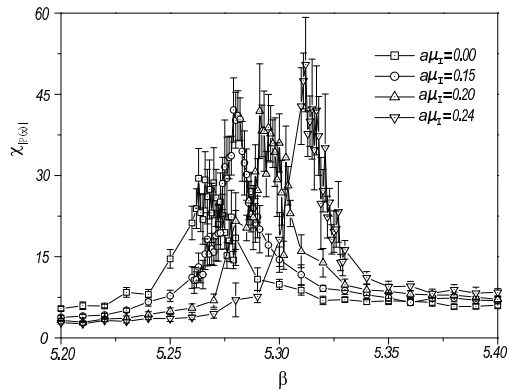}
\end{center}
\caption{Susceptibility of Polyakov loop norm as a function of $\beta$ for
$a\mu_I=0.0,0.15,0.20,0.24$ in $\kappa_1=0.16$.} \label{fig-4}
\end{figure}

\begin{figure} [htbp]
\begin{center}
\includegraphics[totalheight=2.0in]{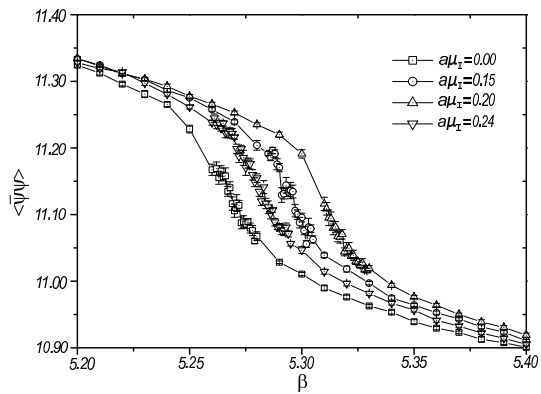}
\end{center}
\caption{Chiral condensate as a function of $\beta$ for $a\mu_I=0.0,0.15,0.20,0.24$ in
$\kappa_1=0.16$ and $\kappa_2=0.08$.} \label{fig-5}
\end{figure}

\begin{figure} [htbp]
\begin{center}
\includegraphics[totalheight=2.0in]{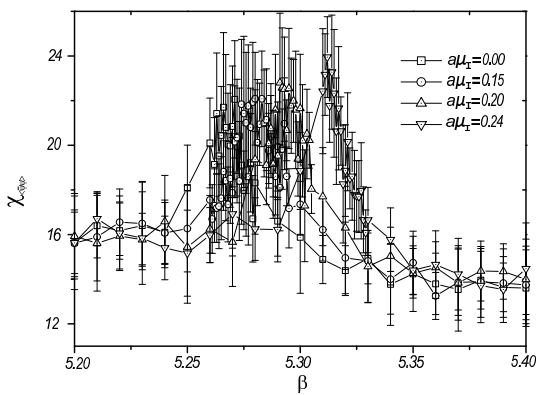}
\end{center}
\caption{Susceptibility of chiral condensate as a function of $\beta$ for
$a\mu_I=0.0,0.15,0.20,0.24$ in $\kappa_1=0.16$ and $\kappa_2=0.08$. } \label{fig-6}
\end{figure}

At First, we fix u,d quark mass parameter $\kappa_1$ at $0.16$, which is far from chiral limit. MC
simulations are done by scanning the parameter $\beta$ for different $a\mu_I$. $2000 \sim 4000$
configurations are measured for every simulation. Polyakov loop $\langle|p(x)|\rangle$ and its
susceptibility $\chi_{|p(x)|}$ as function of $\beta$ are shown in Fig.\ref{fig-3} and
Fig.\ref{fig-4}. As one can see in Fig.\ref{fig-3}, with $\beta$ increased, $\langle|p(x)|\rangle$
raise from approximate zero to obvious non-zero. Especially at $\beta \approx 5.25 \sim 5.35$, the
Polyakov loop show rapidly growth. Exactly at the same location, $\chi_{|p(x)|}$ show sharp peaks
for different $a\mu_I$ in Fig.\ref{fig-4}. The points at the peaks of $\chi_{|p(x)|}$ should be
the critical points of deconfinement phase transition. At the same condition, the chiral
condensate was also studied synchronously. The chiral condensate $\langle\bar\psi \psi\rangle$ and
the susceptibility $\chi_{\bar\psi \psi}$ are illustrated in Fig.\ref{fig-5} and Fig.\ref{fig-6}.
As $\beta$ is increased, $\langle\bar\psi \psi\rangle$ show decrease in Fig.\ref{fig-5}. At the
peaks of $\chi_{|p(x)|}$ in Fig.\ref{fig-4}, $\chi_{\bar\psi \psi}$ show rapid drop and the
susceptibility show sharp peaks too. The results implied that the deconfinement phase transition
maybe coincides with the chiral phase transition. On the other hand, it is obvious in
Fig.\ref{fig-5} that the chiral condensate $\langle\bar\psi \psi\rangle$ is nonzero while QCD is
in the deconfinement phase. It is due to non-preserving the chiral symmetry for Wilson fermion
action . That means the chiral condensate is not any more a good order parameter for the chiral
transition of QCD with Wilson fermions. However, we can also find the chiral critical point by
locating its rapid change and the sharp peak of its fluctuation.

\begin{figure} [htbp]
\begin{center}
\includegraphics[totalheight=2.0in]{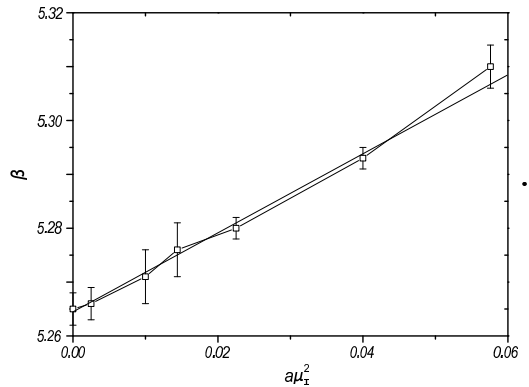}
\end{center}
\caption{The fitted deconfinement phase transition line by collecting the critical points}
\label{fig-7}
\end{figure}

In Fig.\ref{fig-4} and Fig.\ref{fig-6}, the critical points for different $a\mu_I$ can be located
approximately. All critical points can be collected into a critical line $(\beta_C, a\mu_I^C)$ in
$(a\mu_I, \beta)$ plane. For small imaginary chemical potential, this critical line could be
expand into Taylor series. From our simulation results, we obtain the critical line up to the
quadratic term:
\begin{eqnarray}
\label{fits}
  \beta_c = 5.264(1) + 0.75(4)\left( a\mu_I \right)^2 + O\left(a^4\mu_I^4\right).
\end{eqnarray}
The fitted line is shown in Fig.\ref{fig-7}.  By replacing $\mu_I$ with $-i\mu$, the critical line
(\ref{fits}) is transformed from imaginary chemical potential case to real chemical potential
case:
\begin{eqnarray}
             \label{Extropolate}
              \beta_c = 5.264(1) - 0.75(4) \left(a\mu\right)^2 + O\left(a^4\mu^4\right) .
\end{eqnarray}

\begin{figure} [htbp]
\begin{center}
\includegraphics[totalheight=2.0in]{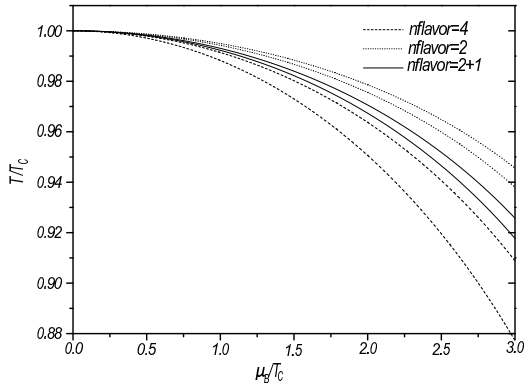}
\end{center}
\caption{Phase diagram on the $(\mu_B, T)$ plane. The area between the two dashed lines is the
error band for 4 flavor Staggered quarks at $m_q=0.05$ from\cite{D'Elia:2002gd}, the dotted lines
is for 2 flavor Staggered quarks at $m_q=0.025$ from \cite{deForcrand:2003hx}, and the solid lines
is for 2+1 flavor Wilson quarks at $\kappa_1=0.16$ and $\kappa_2=0.08$.} \label{fig-8}
\end{figure}

According to the renormalization group relation between the lattice spacing $a$ and $\beta$, the
two loop perturbative expression gives
\begin{eqnarray}
              \label{renormalization}
    a\Lambda_L &= & \exp \bigg(-\frac{4\pi^2}{33-2N_f}\beta \\
    \nonumber  & +& \frac{459-57N_f}{(33-2N_f)^2}\ln \left(\frac{8\pi^2}{33-2N_f}\beta\right) \bigg),
\end{eqnarray}
where $\Lambda_L$ is the lattice QCD scale. The temperature $T$ is related to $a$ and $N_t$ by
$T=1/(aN_t)$. The critical line on the $(\mu,T)$ plane is shown in Fig.\ref{fig-8}.

\begin{figure} [htbp]
\begin{center}
\includegraphics[totalheight=2.0in]{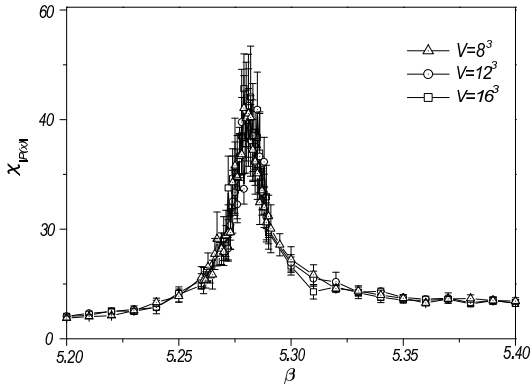}
\end{center}
\caption{Susceptibility of Polyakov loop norm as a function of $\beta$ for different lattice
volume $V=8^3,12^3,16^3$ in $\kappa_1=0.16,\kappa_2 = 0.08$.} \label{fig-9}
\end{figure}

In order to determine the properties of the transition, the dependencies of the transition on the
spatial volume of the lattice are studied. While the lattice volume is $6^3\times 4, 8^3\times 4,
12^3\times 4,$ respectively and $\kappa_1$ is fixed at $0.16$, the results are shown in
Fig.\ref{fig-9}. It is clear that the fluctuation of $\chi_{|p(x)|}$ have no obvious dependence on
the the lattice size and the location of critical points have not shown distinguishable shift. It
suggests that the transition is just crossover while the u,d quark mass is far away from the
chiral limit.

\begin{figure} [htbp]
\begin{center}
\includegraphics[totalheight=2.0in]{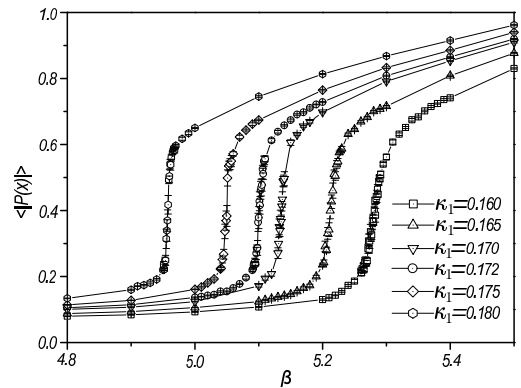}
\end{center}
\caption{Polyakov loop norm as a function of $\beta$ for different $\kappa_1$ in $a\mu_I = 0.15$}
\label{fig-10}.
\end{figure}

\begin{figure} [htbp]
\begin{center}
\includegraphics[totalheight=2.0in]{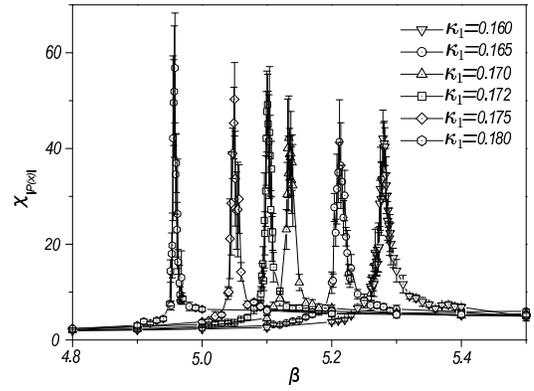}
\end{center}
\caption{Susceptibility of Polyakov loop norm as a function of $\beta$ for different $\kappa_1$ in
$a\mu_I = 0.15$.} \label{fig-11}
\end{figure}

At the following, the dependences of phase transition on the quantity of u,d quark mass near the
chiral limit are studied. Here, the chemical potential $a\mu_I$ is fixed at $0.15$, we scanned u,d
quark mass parameter $\kappa_1$ from $0.16$ to $0.180$. As $\kappa_1$ is increased, which
correspond to decrease the quantity of $u,d$ quark mass, the results of MC simulation at the
critical points show likewise change as above paragraphs. The detailed results of
$\langle|p(x)|\rangle$ and $\chi_{|p(x)|}$ versus $\beta$ were shown in Fig.\ref{fig-10} and
Fig.\ref{fig-11}.

It is obvious in Fig.\ref{fig-10}, for large value of $\kappa_1$ the ascending rate of the
quantity of $\langle|p(x)|\rangle$ near the critical points are more steep and the peaks of
$\chi_{|p(x)|}$ in Fig.\ref{fig-11} at the critical points are more sharp and higher. That implies
that for tiny u,d quark mass the deconfinement phase transition could be of first order.

\begin{figure} [htbp]
\begin{center}
\includegraphics[totalheight=2.0in]{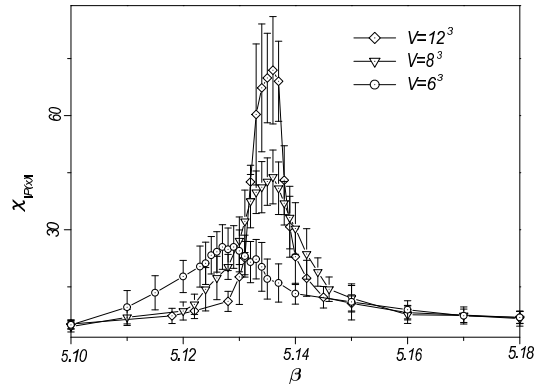}
\end{center}
\caption{Susceptibility of Polyakov loop norm as a function of $\beta$ for lattice volume
$V=8^3,12^3,16^3$ in $\kappa_1=0.170,\kappa_2 = 0.08$ and $a\mu_I=0.150$.}
\label{fig-12}
\end{figure}

The finite volume size effects are checked in $\kappa_1=0.170$ and $a\mu_I=0.150$. $4000\sim
10000$ configurations are measured for every simulation. The $\chi_{|p(x)|}$ as function of
$\beta$ are shown in Fig.\ref{fig-12}. The peak value of $\chi_{|p(x)|}$ is shown higher for
greater lattice volume and the location of the critical point is shown obvious shift with
different lattice volume size. For a second order transition, $\chi_{max}\propto V^\alpha,
\alpha<1$. Our results are fitted to $\chi_{max}\propto V^\alpha$, and we get $\alpha \approx
0.5$. The result suggests that the transition is of second order in the situation.

\begin{figure} [htbp]
\begin{center}
\includegraphics[totalheight=2.0in]{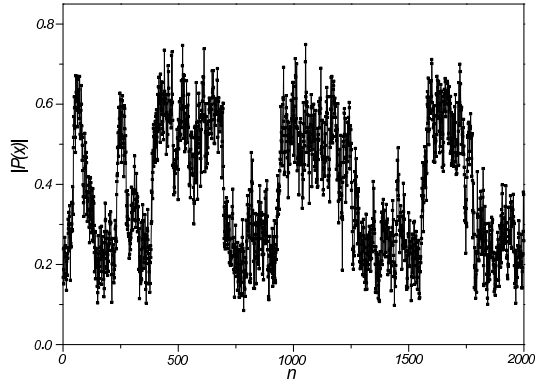}
\end{center}
\caption{History of MC simulation for Polyakov loop norm in $a\mu_I=0.15,\kappa1=0.175$ and
$\kappa_2=0.08$} \label{fig-13}
\end{figure}

\begin{figure} [htbp]
\begin{center}
\includegraphics[totalheight=2.0in]{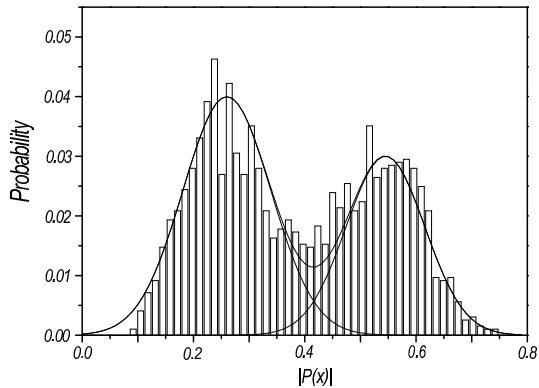}
\end{center}
\caption{Histogram of Polyakov loop in $a\mu_I=0.15,\kappa1=0.175$ and $\kappa_2=0.08$, the solid
curves are fitted Gaussian curve.} \label{fig-14}
\end{figure}

As one can see in Fig.\ref{fig-10}, for $\kappa_1>0.175$ the variety of $\langle|p(x)|\rangle$ at
the critical points is very rapid. In $\kappa_1=0.175$ and $\beta=5.10$, the history of MC
simulation is shown in Fig.\ref{fig-13} and the histogram of the Polyakov loop norm is shown in
Fig.\ref{fig-14}. Two plateaus at $|p(x)|\approx 0.25$ and $|p(x)|\approx 0.55$ alternately
appeared in the history of MC simulation and two peaks respectively at $|p(x)|\approx 0.25$ and
$|p(x)|\approx 0.55$ are distinguishable in Fig.\ref{fig-14}. Especially for tiny $u,d$ quark mass
, the structure of two peaks in the histogram are shown more outstanding. The results suggest,
while the quantity of $u,d$ quark mass is decreased to be close to the chiral limit and the s
quark mass is large enough, the deconfinement phase transition should be of first order.

\begin{figure} [htbp]
\begin{center}
\includegraphics[totalheight=2.0in]{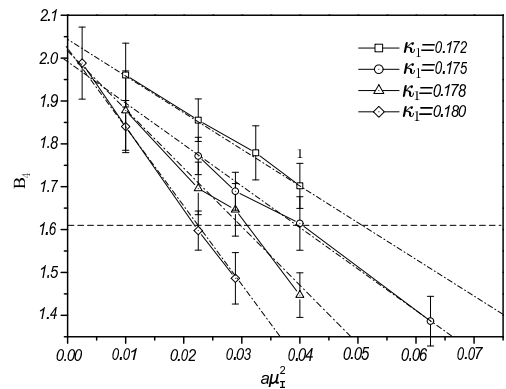}
\end{center}
\caption{The Binder cumulant $B_4$ of Polyakov loop norm for different $(a\mu_I)^2$. The dashed
line corresponds the critical line $B_4=1.604$. The dash\_dotted line are the fitted lines for
different $\kappa_1$} \label{fig-15}
\end{figure}

\begin{figure} [htbp]
\begin{center}
\includegraphics[totalheight=2.0in]{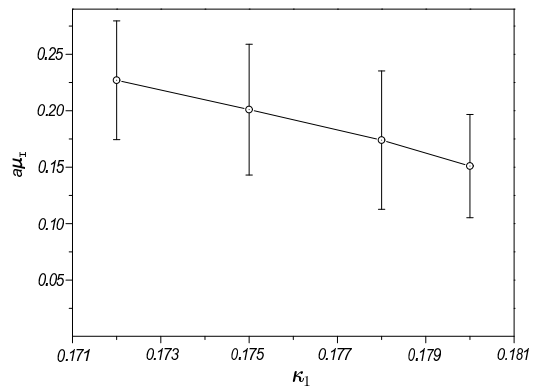}
\end{center}
\caption{The second order critical points in the $(a\mu_I,\kappa_1)$ plane.} \label{fig-16}
\end{figure}

In order to investigate the universality class of the phase transition, Binder cumulants are
calculated in different quark mass and chemical potential. For $\kappa_2=0.08$, MC simulation are
performed by scanning $a\mu_I$. $2000\sim 8000$ configurations are measured for every simulation.
The results of B4 are shown in Fig.\ref{fig-15}. With increasing $a\mu_I$, B4 drops. For
$a\mu_I\rightarrow 0$, $B4 \approx 2.0$. A second order point, which is chosen as the critical
value $B4 = 1.604$ of 3D Ising model, separate the first order from the crossover. The phase
transition for $B4 < 1.604$ should be of first order and for $B4 > 1.604$ could be crossover.
According to above results, the phase transition near $B4 \approx 1.604$ should be of second order
and for $B4 \gg 1.604$ should be crossover, their exact border can not be located by traditional
methods. From Fig.\ref{fig-15}, we got the second order critical cure in Fig.\ref{fig-16} in the
$(a\mu_I,\kappa_1)$ plane. For small $a\mu_I$, we can Taylor-expand the critical line in
Fig.\ref{fig-16} as:
$$
 \kappa_1^c(a\mu_I) = \kappa_1^c(a\mu_I=0) + b (a\mu_I)^2+O(a^4{\mu_I}^4).
$$
From Fig.\ref{fig-16}, we get $\kappa_1^c(0)=0.1864(4),b=-0.280(3)$. By replacing $\mu_I$ by
$-i\mu$, the critical line is continued from imaginary chemical potential to real chemical
potential:
\begin{eqnarray}
\label{Taylor-expansion}
 \kappa_1^c(a\mu) = 0.1864(4) + 0.280(3)(a\mu)^2+O(a^4{\mu}^4).
\end{eqnarray}
According to (\ref{Taylor-expansion}), for small $\kappa_1$ (large u,d quark mass) the critical
point A in Fig.\ref{fig-1} is more close to the T axis. Especially while $\kappa_1=\kappa_1^c(0)$,
the critical point A will collide with T axis and disappear.  The result coincides with
\cite{deForcrand:2006pv}.

\section{Conclusions}
In this paper, we have used MC simulation on lattice with 2+1 flavors Wilson fermions to determine
the structure of the phase diagram of QCD at finite temperature and small density. As we have
discussed in the part of instruction, the operation that the fermionic determinant is token the
fourth root to thin the species doubling problem in MC simulation with staggered fermions can
arise the locality problem in simulations. Whereas the MC simulations with Wilson fermions on
lattice have no this puzzled problem.

\begin{figure} [htbp]
\begin{center}
\includegraphics[totalheight=2.0in]{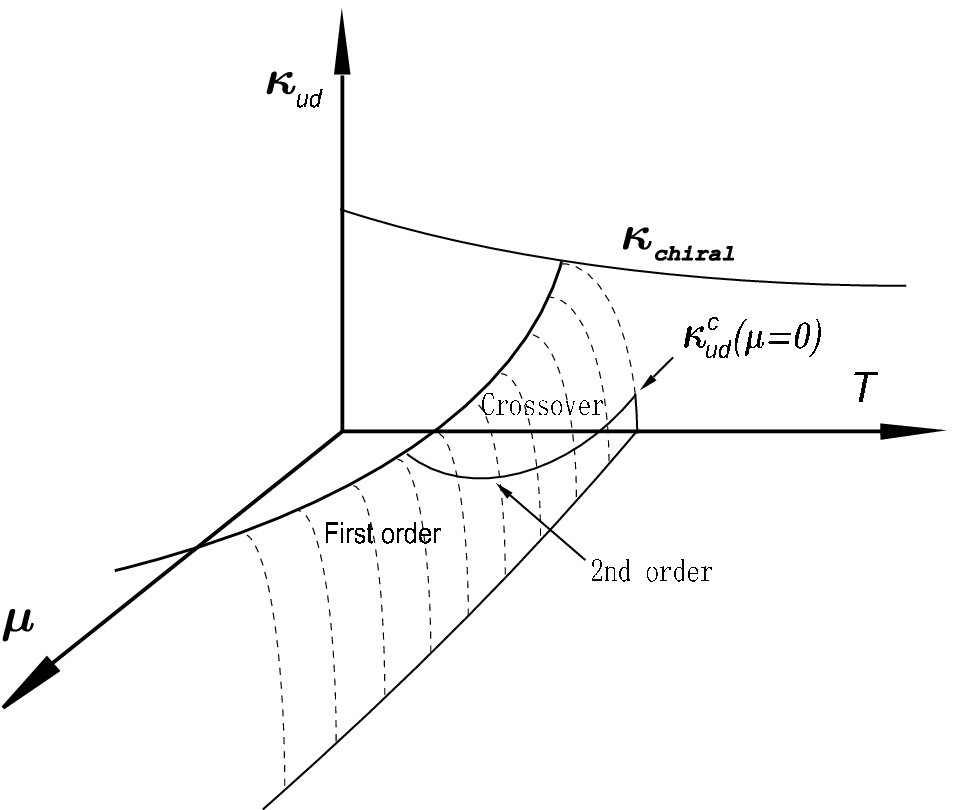}
\end{center}
\caption{The schematic phase diagram of QCD with 2+1 flavors of Wilson quarks in the $(\mu, T,
\kappa_{ud})$ space. A critical line separate the phase transition interface into the first order
and the crossover.  The phase transition is of second order near the critical line.}
\label{fig-17}
\end{figure}

According to our results of MC simulation, the schematic expected phase diagram was shown in
Fig.\ref{fig-17}. Below the chiral limit surface $\kappa_{ud}=\kappa_{chiral}$ and above the
surface $\kappa_{ud}= 0$, the space was separated into two parts by the phase transition
interface. QCD is in hadronic matter phase inside the phase transition interface and in QGP phase
outside the phase transition interface. The phase transition interface is separate into the first
order and the crossover by a second order critical curve. The critical curve is just the critical
point A in Fig.\ref{fig-1} which is function of quark mass.

\section{Acknowledgments}
 We thank M. D¡¯Elia, C. DeTar, P. de Forcrand, E. Gregory, and M. Lombardo for useful discussions. All
simulations are performed on the HPCC of Linux PC cluster in SYSU and YZU. At last, I am very
sorrowful to hear the death of my cooperator, Prof.Xiangqian.Luo.

\end{document}